\documentclass[twoside,11pt]{article}

\usepackage{jmlr2e}
\usepackage{subcaption}
\usepackage{wrapfig}

\jmlrheading{1}{2017}{1-48}{11/17}{10/00}{Mark Hamilton, MMLSpark Team}

\ShortHeadings{Flexible and Scalable Deep Learning with MMLSpark}{Hamilton}
\firstpageno{1}

\begin{document}

\title{Flexible and Scalable Deep Learning with MMLSpark}

\author{\name Mark Hamilton \email marhamil@microsoft.com \\
		\name Sudarshan Raghunathan \\
		\name Akshaya Annavajhala\footnotemark  ,
        \name Danil Kirsanov\footnotemark[\value{footnote}] ,
        \name Eduardo de Leon\footnotemark[\value{footnote}] ,
        \name Eli Barzilay\footnotemark[\value{footnote}] ,
        \name Ilya Matiach\footnotemark[\value{footnote}] ,
        \name Joe Davison\footnotemark[\value{footnote}] ,
        \name Maureen Busch\footnotemark[\value{footnote}] ,
        \name Miruna Oprescu\footnotemark[\value{footnote}] ,
        \name Ratan Sur\footnotemark[\value{footnote}] ,
        \name Roope Astala\footnotemark[\value{footnote}] ,
        \name Tong Wen\footnotemark[\value{footnote}] ,
        \name ChangYoung Park\footnotemark[\value{footnote}]  \\
       \addr Azure Machine Learning\\
       Microsoft New England Research and Development\\
       Boston, MA 02139, USA}
\editor{}
\footnotetext[\value{footnote}]{All authors significantly contributed to the library, remaining names in alphabetical order. For more detail see \url{aka.ms/mmlspark}}

\maketitle

\begin{abstract}
In this work we detail a novel open source library, called MMLSpark, that combines the flexible deep learning library Cognitive Toolkit, with the distributed computing framework Apache Spark. To achieve this, we have contributed Java Language bindings to the Cognitive Toolkit, and added several new components to the Spark ecosystem. In addition, we also integrate the popular image processing library OpenCV with Spark, and present a tool for the automated generation of PySpark wrappers from any SparkML estimator and use this tool to expose all work to the PySpark ecosystem. Finally, we provide a large library of tools for working and developing within the Spark ecosystem. We apply this work to the automated classification of Snow Leopards from camera trap images, and provide an end to end solution for the non-profit conservation organization, the Snow Leopard Trust.
\end{abstract}


\section{Introduction}

Deep learning has recently flourished in popularity due to several key factors: superb performance in a variety of domains, quick training time due to the rich counter-factual information contained in the gradient, low memory training footprint of stochastic gradient descent, and the proliferation of automatic differentiation frameworks that allow users to easily define a large space of models without implementing gradient computations. As datasets grow in size, and models become larger and more complex, it becomes increasingly necessary to develop methods to easily and efficiently parallelize the training and evaluation of these models. 

In the space of high performance parallel computing, Apache Spark has recently become one of the industry favorites and standards. Its success can be attributed partly to its rapid in-memory architecture, its guaranteed fault tolerance, and its high level functional APIs for machine learning, graph processing, and high-throughput streaming. However, Spark, and its machine learning library, SparkML, provide almost no support for deep learning and existing models such as SparkML's word2vec and logistic regression do not leverage modern computational architectures such as Graphics Processing Units (GPUs) or Automatic Differentiation. In this work we aim to fill this gap by integrating the performant and flexible deep learning framework, The Microsoft Cognitive Toolkit (formerly CNTK).

In this work we first provide background on deep learning with the Cognitive Toolkit (section \ref{sec:CNTK}), Apache Spark (section \ref{sec:Spark}), and its machine learning library, SparkML (section \ref{sec:SparkML}). We then outline the contributions of MMLSpark and documenting the work required to integrate the two frameworks. Section \ref{sec:CNTKJava} documents Java language bindings for the toolkit, and section \ref{sec:CNTKSpark} describes the integration with SparkML, and our performance optimizations (section \ref{sec:Opt}). Next, we briefly overview our integration of OpenCV (section \ref{sec:OpenCV}), and our tool to automatically integrate this work into PySpark and SparklyR (section \ref{sec:PySpark}). In section \ref{sec:Perf} we investigate the scalability of this method. In section \ref{sec:Leopard} we document our work with the Snow Leopard Trust, and demonstrate a large increases in accuracy and speed over traditional ML methods for automated Snow Leopard identification from remote camera trap images.

\section{Background}
\subsection{Deep Learning with the Cognitive Toolkit}
\label{sec:CNTK}

Deep learning has steadily gained popularity throughout the past several years due to a variety of complementary reasons. First and foremost, deep learning has shown incredible performance in almost all domains it has been applied to from speech processing \cite{dlspeech}, translation \cite{dltranslation}, image classification \cite{dlimage}, and probabilistic modeling \cite{dlprob}. This stellar performance has arose in part to significant advances in network optimization, primarily arising from advances in gradient based methods and computations. Networks often take the form of differentiable functions with latent parameters, optimized using empirical risk minimization. This gradient acts as rich source of counter-factual information, effectively allowing the optimizer to explore and understand the local neighborhood of a network's parameter space with only a single function and gradient computation. In recent deep learning frameworks such as the Cognitive Toolkit, Tensorflow, and Caffe, the calculation of the gradient has been abstracted away. This significantly simplifies code and simultaneously allows for automatic compilation of gradient computations to GPU architectures. It is now easier than ever to develop performant and novel network architectures using these automatic differentiation libraries. 

The Cognitive Toolkit (CNTK) is an open source deep learning framework created and maintained by Microsoft. It is written in C++, but has ``bindings'' or interfaces in Python, C$\sharp$, and a domain specific language called BrainScript. In the Cognitive Toolkit, users create their network using a lazy, symbolic language. This symbolic representation of the network, and its gradients, can be compiled to performant machine code, on either a CPU or a GPU. Users can then execute and update parameters of this symbolic computation graph, usually using gradient based methods. The toolkit differentiates itself from other deep learning frameworks as it supports dynamic networks through sequence packing \cite{cntkseq}, and model compression for low latency operations. We have chosen the Cognitive Toolkit as our first deep learning framework to integrate with Spark because of its superior performance relative to other deep learning frameworks, and its first class treatment of parallel processing.  Furthermore, the Cognitive Toolkit has well formed abstractions for training and reading data from a variety of sources that we plan to leverage in future optimizations of the work.

\subsection{Apache Spark}
\label{sec:Spark}

Apache Spark is an open source distributed computing platform that generalizes and combines map-reduce and SQL style parallelism into a single a functional language. Apache spark vastly outperforms conventional map-reduce algorithms on platforms like Hadoop, because of its heavy usage of in-memory computing, and its sophisticated Catalyst query optimizer \cite{sparksql}. Like CNTK, Spark exposes its core functionality to other languages using bindings. Currently, Spark has bindings in Python (PySpark), Java, R (SparklyR), C$\sharp$ and F$\sharp$ (Mobius). Users of Spark construct lazy, symbolic representations of their parallel jobs using a functional language built in Scala. This symbolic representation can then be optimized and compiled to custom parallel code that minimizes inter-node communication and time intensive IO operations. Spark operations are completely fault-tolerant and large numbers of nodes can be killed without effecting the overall status of the job. Furthermore, Spark clusters support elastic workloads with a feature called ``Dynamic Allocation''. In this mode, nodes can be added or removed from a job depending on cluster utilization. This allows for computations to scale adaptively to take full advantage of a cluster if necessary or to free up costly unused resources.

A central abstraction in the Spark ecosystem is the ``DataFrame'', or a distributed table of ``Rows''. Rows are type safe containers that can hold various forms of data. Each DataFrame has a pre-defined schema, or collection of meta-data that can be used to verify the correctness of a computation before execution, similar to a type system. Roughly, each node of the cluster contains a small subset, or ``partition'' of a DataFrame's rows, and computations are performed on each machine's subset of the data. For more complex computations, Spark can perform ``shuffles'' or transfers of data between nodes. Furthermore, within the same DataFrame API, Spark supports high throughput streaming work-flows, allowing for computations on thousands of records per second with sub-second latencies. 

\subsection{SparkML}
\label{sec:SparkML}

SparkML is a library built on top of Spark's DataFrame API that gives users access to a large number of machine learning, and data processing algorithms. SparkML is similar to other large machine learning libraries such as python's SciKit-learn, but is significantly more flexible. For example unlike SciKit-learn, SparkML supports multiple typed columns and has a rich type system allowing for static validation of code. Furthermore, it can scale almost indefinitely. 

SparkML's central abstraction is a PipelineStage, or a self contained unit of a data science workload that can be composed with other units to build large custom Pipelines for a given machine learning task. This central class is split into two subclasses: Estimators, and Transformers. Estimators abstract the components of the machine learning work-flow that learn, or extract state, from a dataset. Estimators can be fit to a DataFrame to produce a Transformer. A Transformer is a referentially transparent function that can transform a DataFrame into another DataFrame, often using the state learned from the fitting process. This neatly separates the tasks of machine learning into training and evaluation and can be useful to ensure that one is not learning from a testing dataset. As a concrete example, one can consider PCA as an Estimator that finds the principal components using a training dataset, and returns a PCA transformer, or an object that can transform a testing dataset into the space whose basis is the principal component vectors learned from the training dataset. 

SparkML supports conventional classifiers and regressors such as Logistic Regression (LR), Support Vector Machines (SVMs), Boosted Decision Trees (BDTs), and Generalized Linear Models (GLMs). Spark also supports unsupervised methods such as clustering, principal component analysis (PCA), Latent Dirichlet Allocation (LDA), word2vec, and TF-IDF weighting. However, SparkML currently does not support deep networks, or Kernel methods, which poses a significant challenge to those looking to use the powerful methods from these fields.

\section{Methods}
\subsection{The Cognitive Toolkit in Java}
\label{sec:CNTKJava}

In order to integrate the Cognitive Toolkit and Spark, we first need them to ``speak the same language''. One way to accomplish this is to use CNTK's python bindings and Spark's python bindings called PySpark. However, this approach suffers from the unnecessary baggage of the python interpreter, which can considerably slow operations as input and output data needs to be transferred from the JVM to the python interpreter, and the compute context needs to be switched from Scala to python and back again after execution. 

To avoid the overhead of the python interpreter, we focused on bringing CNTK to Scala by creating Java language bindings for CNTK. To bring CNTK into Java, we used the Simple Wrapper and Interface Generator (SWIG) \cite{swig}. This tool exposes C++ code in a wide variety of languages such as Java, C$\sharp$, OCaml, python and many more. SWIG first inspects a C++ file, and translates each programming construct to an equivalent construct in the target language. For Java, each C++ function translates to a corresponding Java function, and each C++ class translates to a Java class. This mapping is controlled by a specification called an ``interface file''. In this file one can define custom maps from one type system to another, inject custom code for usability of the generated bindings, and programmatically control translations with a basic macro system. To translate C++ to Java, many of the C++ functions are wrapped using the Java Native Interface (JNI), a language feature that allows Java code to call into native C and C++ code efficiently. This results in performant C++ code that has look and feel of a normal Java function. The default bindings SWIG produces are generally not very elegant, as complex language features such as generics are lost in translation. To mitigate this we have contributed a custom interface file that makes these Java bindings as idiomatic as possible and have contributed this binding generation process back to the Cognitive Toolkit. We also publish a Java Archive (JAR) of the bindings, complete with automatic loading of native libraries, to Maven Central for every CNTK release. This is the first time CNTK has been brought into a virtualized language, and will enable deep learning on a large class of new devices running on the JVM. Furthermore, this work enables research at the intersection of type theory and deep learning, allowing for the possibility of richer abstractions when working with CNTK computation graphs. 

\subsection{The Cognitive Toolkit in Spark}
\label{sec:CNTKSpark}

To bring our CNTK Java bindings into Spark's Scala ecosystem, we relied on Scala's seamless inter-operation with Java. More specifically, every class in Java is treated as a type in Scala, and Scala preserves all of Java's complex language features. Effectively, we get CNTK Scala bindings for free! We hope that other languages on the JVM will be able to leverage these bindings similarly. This brings CNTK to Scala in an idiomatic and efficient way. 
%
%

To bring CNTK to Spark, we leverage Spark's ability to distribute and execute custom Scala code inside of a mapping operation. More specifically, we use the ``mapPartitions'' function that enables parallel evaluation of a custom Scala function on each partition. This is strictly more general than Spark's ``map'' operation as the function can depend on \textit{all} of the rows in a partition. It is common to use this pattern to move slow, setup-type, operations outside of the main loop of the map to improve performance. Using MapPartitions we can take advantage of CNTK's Same Instruction, Multiple Data (SIMD) optimizations that perform computations on a large number of data concurrently, potentially on a GPU. This speeds performance significantly, on both the CPU and GPU. 

We built on this MapPartitions operation and create a SparkML transformer, called the ``CNTKModel'', that distributes, loads, and executes arbitrary CNTK computation graphs inside of SparkML pipelines. We expose parameters that give users control over the CNTK model object, mini-batch size, and evaluation of subgraphs of the CNTK graph. The ultimate is useful for transfer learning, where one is interested in inner activations of a network \cite{deeptransfer}

\subsection{Optimizations}
\label{sec:Opt}

In order to improve performance in streaming and repeated evaluation scenarios we add several memory and performance enhancing optimizations. Initially, we loaded the CNTK model on each partition, and evaluated the model on every transformation. In Spark, each node can hold multiple partitions, so this wastes time and memory on loading the model multiple times on each node. To improve this, we used Spark's system for broadcasting data from the driver node to each of the worker nodes using bit-torrent communication to minimize overhead. 

%
%

\subsection{Multi-GPU Training}
In addition to parallel evaluation of deep networks, we also provide several solutions for rapid training of CNTK models. CPU Spark clusters excel at large scale extract, transform, and load (ETL) operations. However these clusters lack GPU acceleration and InfiniBand support, which degrades the performance of high communication network training. As a result, we have investigated hybrid systems consisting of dedicated multi-gpu machines, tethered to the Spark cluster over a virtual network. These systems use several CNTK distributed training modes such as the highly optimized 1-bit SGD \cite{1bit}. We provide Azure Resource Manager templates so that users can deploy these solutions.

\subsection{Integrating OpenCV}
\label{sec:OpenCV}

SparkML also lacks support for complex datatypes such as images. We have worked with the Spark community to contribute a schema for image data-types, and a collection of readers for ingesting and streaming images and binary files to the core Spark library. Building on this work, we have integrated the popular image processing library OpenCV \cite{opencv} as a SparkML transformer. Like the CNTKModel, the OpenCV transformer uses Java bindings for OpenCV that call into fast native C++ code. Our OpenCV transformer allows us to perform most OpenCV operations in parallel on all nodes of the cluster. Furthermore integration automatically chains all evaluations of native code so that data does not need to be Marshalled to and from the JVM between adjacent calls to OpenCV.
%
%

\subsection{Automated Generation of PySpark}
\label{sec:PySpark}

\begin{figure}[h!]
  \centering
  \begin{subfigure}[b]{0.4\linewidth}
    \includegraphics[width=\linewidth]{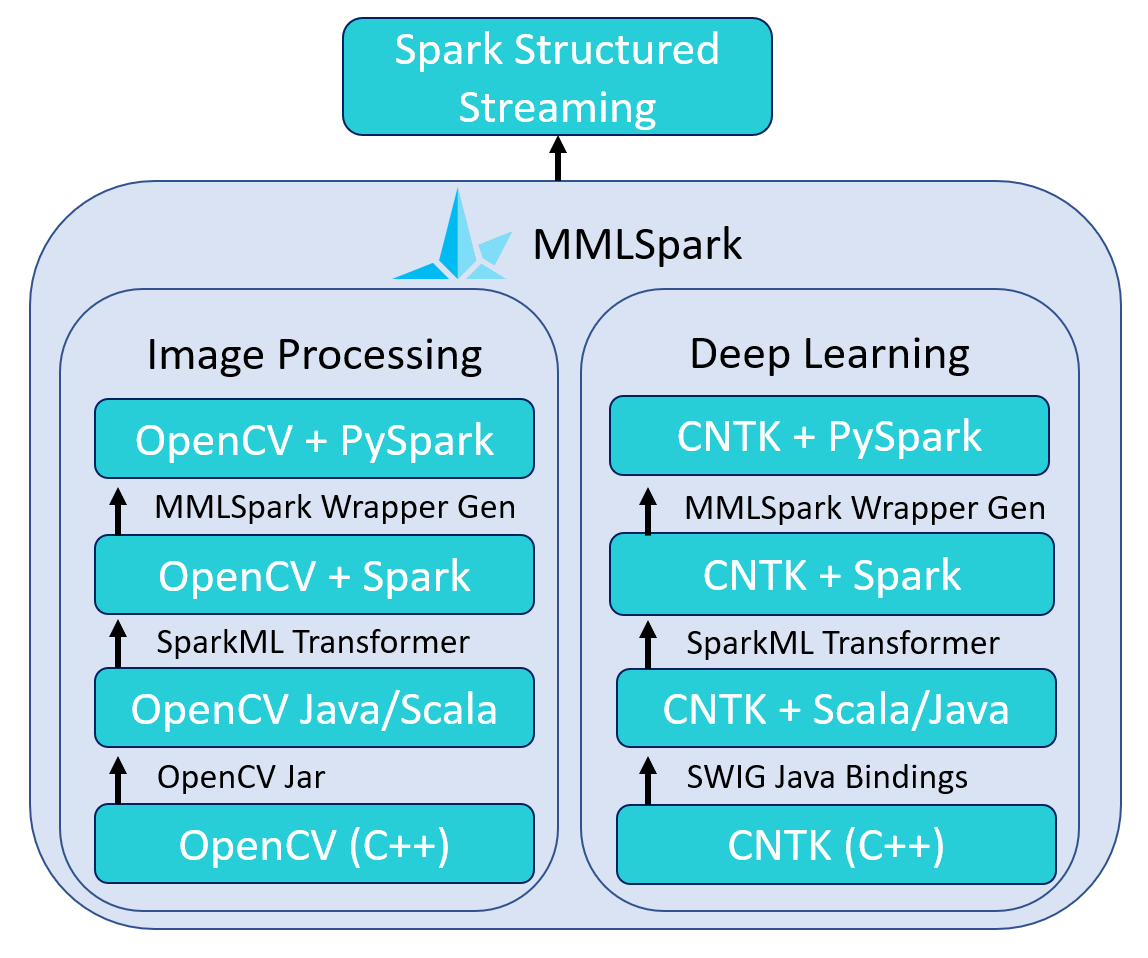}
  \end{subfigure}
  \begin{subfigure}[b]{0.4\linewidth}
    \includegraphics[width=\linewidth]{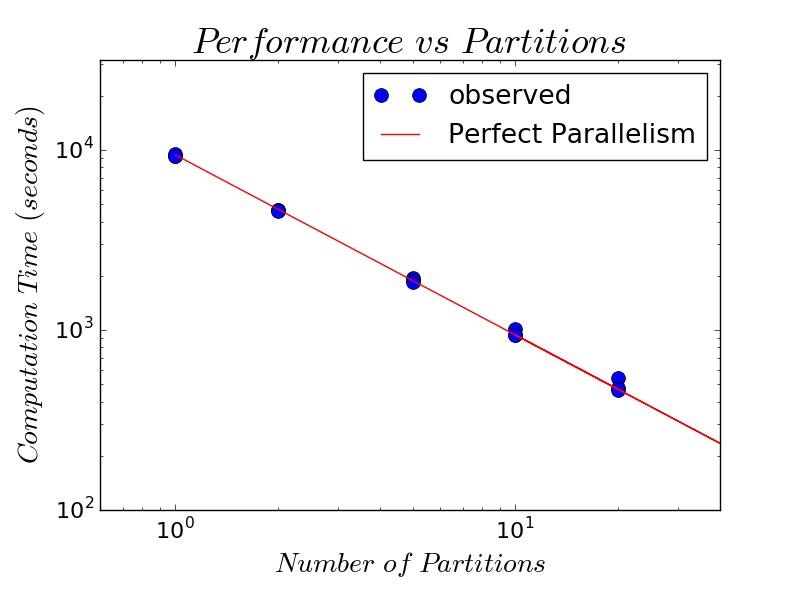}
  \end{subfigure}
  \caption{Left: Architectural overview of some salient features of MMLSpark. Right: Performance of the ``ImageFeaturizer'' transformer that contains an OpenCV image resize, and deep featurization with ResNet-50}
  \label{fig:arch}
\end{figure}

Spark supports a wide variety of ecosystems through language bindings in Python, R, Java, C$\sharp$ and F$\sharp$. However, wrapper code for these libraries is mainly written ``by hand''. This a difficult task involving detailed knowledge of both the source and target ecosystems, and their communication platforms. Furthermore, this results in a large amount of boilerplate code, that impedes adding estimators or transformers to the spark ecosystem. To eliminate the need to manually create and maintain these language bindings, we have created an automated wrapper generation system. The system loads and inspects the types and parameters of any transformer or estimator using Scala reflection. It then generate a usable and idiomatic python wrapper for the class that can be used in PySpark. We also use this tool to create language bindings for the SparklyR ecosystem. We expose all of our estimators and transformers to these languages, proving the generality and reliability of this system. A schematic overview of the process is shown in Figure \ref{fig:arch}.

\subsection{Additional Contributions}
\label{sec:Contrib}

In addition to the aforementioned contributions to the Spark and CNTK libraries, we provide the following additional contributions to SparkML: 
\begin{itemize}
	\itemsep-.3em 
	\item{An estimator for automatic featurization, so that users can immediately begin experimenting with learning algorithms.}
    \item{An infrastructure for the automated testing of SparkML estimators for serialization, fuzzing, python wrapper behavior, and good design practices.}
    \item{A single estimator for high performance text featurization.}
    \item{A transformer for deep network transfer learning on images.}
    \item{A high availability content delivery network of state-of-the-art pre-trained deep networks. This delivery system has a Scala and Python API for querying and downloading with caching and checksumming.}
    \item{Several new pipeline stages for performing Spark SQL operations, repartitioning, caching. These help close many gaps in the SparkML pipeline AP.}
    \item{A high performance vector assembler that considerably outperforms SparkML's.}
    \item{A series of Jupyter notebooks documenting and demonstrating our contributions on real datasets.}
    \item{An estimator for image dataset augmentation with OpenCV transformations.}
\end{itemize}

\section{Performance and Scalability}
\label{sec:Perf}

We test our experiment with on a dataset $82,382$ real images of various sizes (upwards of $400 \times 600$ pixels) encoded as JPEGs. The experiment reads data from ``WASB'' storage, a HDFS client backed by Azure Blob storage, then passes them through our ImageFeaturizer transformer. The ImageFeaturizer internally pipelines the data into OpenCV to scale it down to the proper size and unroll the scaled image into a vector. The transformer then feeds the result through ResNet-50, a 50 layer deep state of the art convolutional network with recurrent connections. The computation time as a function of the number of executors in the spark cluster is shown in Figure \ref{fig:arch}. We have tested this pipeline with up to 20 executors, but have observed similar scaling behavior beyond that point. As a rule of thumb, scaling behavior tends to improve with larger datasets. 
%
%
%
%

\section{Experimental Validation}
\label{sec:Leopard}

We have collaborated with the non-profit organization The Snow Leopard Trust to help them develop an automated system for organizing their extremely large database of camera trap images. This is not only a novel scientific experiment, but also validates the library and ensures that our abstractions work well in practice. 

\subsection{Snow Leopard Conservation}

Snow leopards are a rare species of large cat found throughout the mountains of northern and central Asia. These large cats are one of the top predators of this ecosystem, but are incredibly rare with only $3,000-7,000$ cats spread out across 1.5 million square kilometers. These creatures are endangered, and are constantly threatened by loss of habitat through mining and climate change, poachers, and retribution killing. Despite their endangered status, relatively little is known about snow leopard behavior, ecology, population numbers, and population trend. These data are incredibly important for understanding the risk of snow leopard extinction and where to target conservation efforts. Furthermore, robust habitat information is necessary in order to lobby for legal protection of habitat.

\begin{wrapfigure}{l}{0.5\textwidth}
  \centering
   \includegraphics[width=.48\textwidth]{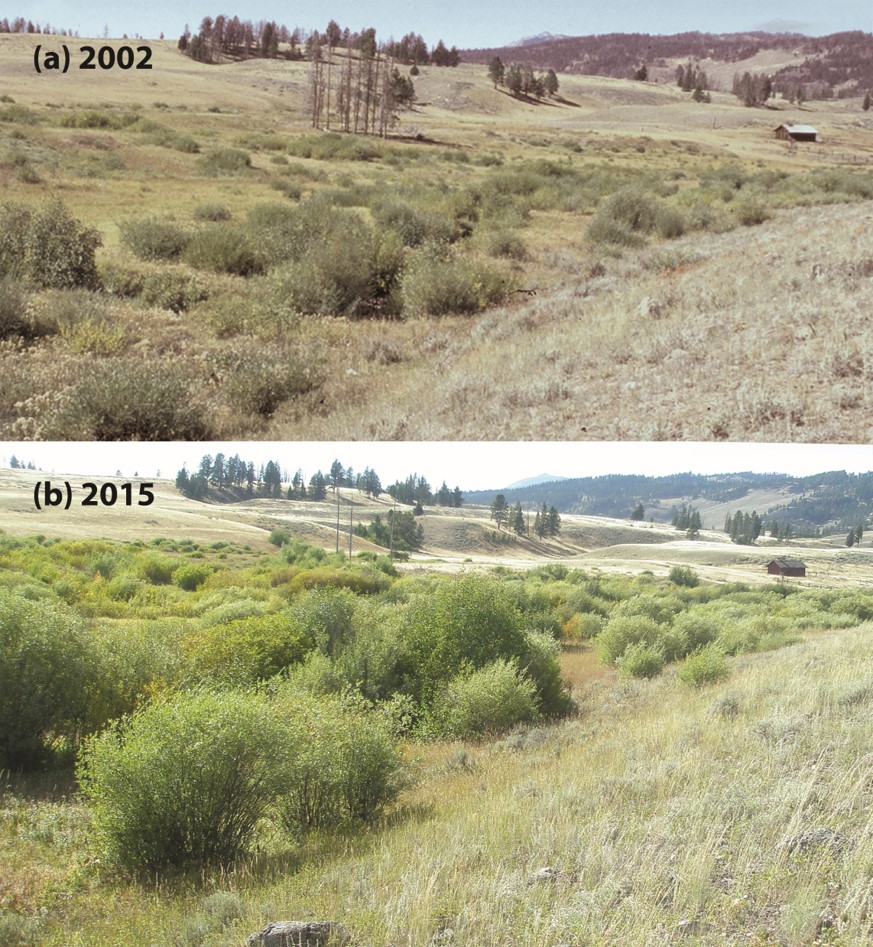}
  \caption{A comparison of vegetation in Yellowstone National Park before (above) and after (below) the reintroduction of the Grey wolf. Figure originally from \cite{yellowstone}}
  \label{fig:yellowstone}
\end{wrapfigure}

Snow leopards occupy a critical niche in the Himalayan ecosystem as one of the apex predators. Apex predators have been shown to help stabilize ecosystems, by naturally controlling populations of secondary lifeforms in the ecosystem. This allows for primary life forms, such as plant life, to thrive and actively improves soil stability. Added vegetation can introduce several new niches, increasing biodiversity and ecosystem stability. Stunning evidence of this effect can be seen in Figure \ref{fig:yellowstone}, which shows the effect of re-introducing the Grey wolf to the Yellowstone ecosystem in the early 2000's. The health and stability of the Himalayan ecosystem is critically important as it provides water for for over 1.4 billion people \cite{watershed}.

Surveying snow leopards is an extraordinarily difficult task. This task is made difficult because leopards are incredibly rare, live in remote and difficult to reach areas, and are mainly nocturnal. As a result, the Snow Leopard Trust uses remote camera traps to survey snow leopard locations over 1,700km$^2$ of potential habitat. These camera traps are equipped with regular and night vision settings, and take photographs when an attached motion sensor is triggered. These cameras generate a large number of images that need to be manually sorted to find the leopards. For many images, it is difficult for a human to identify hidden leopards, as they are often partially obscured and highly camouflaged. This task can take up to 20 thousand man hours per each million images. This task is insurmountable for most small nonprofits like the Snow Leopard Trust. Hence, it is critically important to automate this task.  

\subsection{Dataset}

We use a dataset from Snow Leopard Trust that contains roughly $8,000$ labeled images of snow leopards and non snow leopards as well as roughly $300,000$ unlabeled images. These images are from their most recent set of surveys in the last 5 years. The images are primarily black and white night vision and optical images. A small number $~5\%$ of images are in full color. Only around $10\%$ of the images are Snow Leopards. The majority ($90\%$) of the images are of grass, goats, foxes, and occasionally people. We split the labeled dataset into a training ($80\%$) and testing ($20\%$) set based on the cameras used in the analysis. Splitting the cameras into training and test sets is necessary, as subsequent frames in a camera burst are highly correlated. If one splits a single burst into both sets, information will leak from the training set, and the final testing performance will be overestimated. We did not have to ``clean'' the dataset, or remove any images, as anything that did not contain a snow leopard was marked as a negative image. At the time of publication, we use MMLSpark version 0.10, Spark version 2.2 and CNTK version 2.3.

\begin{figure}[h!]
  \centering
  \includegraphics[width=\linewidth]{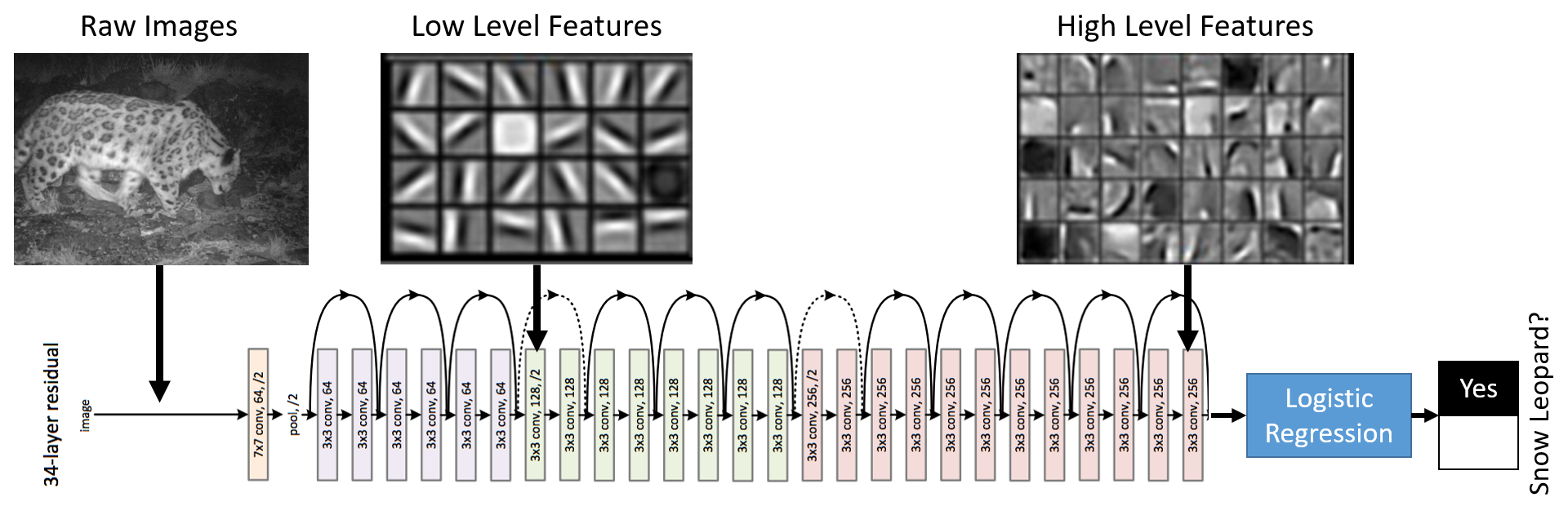}
  \caption{Schematic overview of basic transfer learning algorithm. We use ResNet50 and truncate the last layer(s). We then append a SparkML logistic regression estimator and train the model using the features from the preceding network. Feature visualizations originate from a similar convolution architecture by \cite{filters}.}
  \label{fig:network}
\end{figure}

\subsection{Model and Results}

To learn an automated classification system, we leverage deep transfer learning to learn a classifier that can successfully identify snow leopards with high precision and recall. We improve the classifier through the use of dataset augmentation and ensembling leopard probabilities over bursts of images. The simplest model uses features from ResNet50, trained on the ImageNet corpus. ResNet50 is a state of the art network that combines convolutional and recurrent components to capitalize on the spatial regularity of natural images and effectively learn the optimal depth of the network \cite{resnet}. We truncate the last, image-net specific layer(s) but keep a majority of the early layers. This lets us leverage the domain-general low to mid level features and adapt the network to the new classification task. To adapt the network, we append and train a SparkML logistic regressor that maps the features to a binary label. This effectively adds an additional fully connected layer to the truncated network. A schematic overview of the algorithm is provided in Figure \ref{fig:network}. We refer to these algorithms as ``RN1'' and ``RN2'' for ResNet50 with 1 and 2 truncated layers respectively.

To improve performance of the algorithm, we augment the dataset using MMLSpark's ``ImageSetAugmenter'' transformer. This Transformer enriches the dataset by flipping the images about the horizontal axis. The ImageSetAugmenter doubles the training data, which improves the learned classifier. At test time, the scores for both image parities are averaged together. This reduces the variance of the final estimate and improves performance. We refer to this algorithm as ``RN2+A'' for the base algorithm with dataset augmentation.

\begin{wrapfigure}{l}{0.4\textwidth}
  \centering
   \includegraphics[width=.38\textwidth]{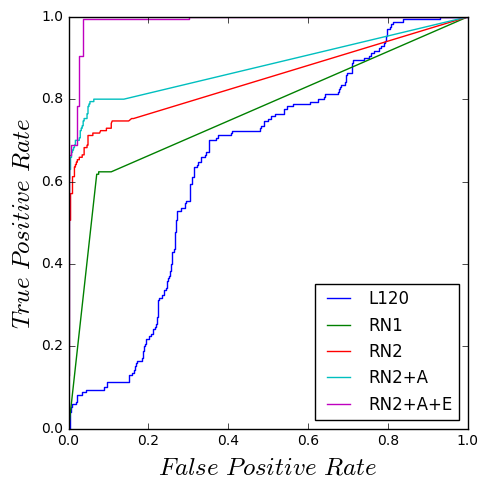}
  \caption{Receiver Operator Characteristic (ROC) curves for each algorithm. Note the superior performance of ``RN2+A+E'' to all other methods tested}
  \label{fig:roc}
\end{wrapfigure}

\begin{figure}[h!]
  \centering
  \includegraphics[width=\linewidth]{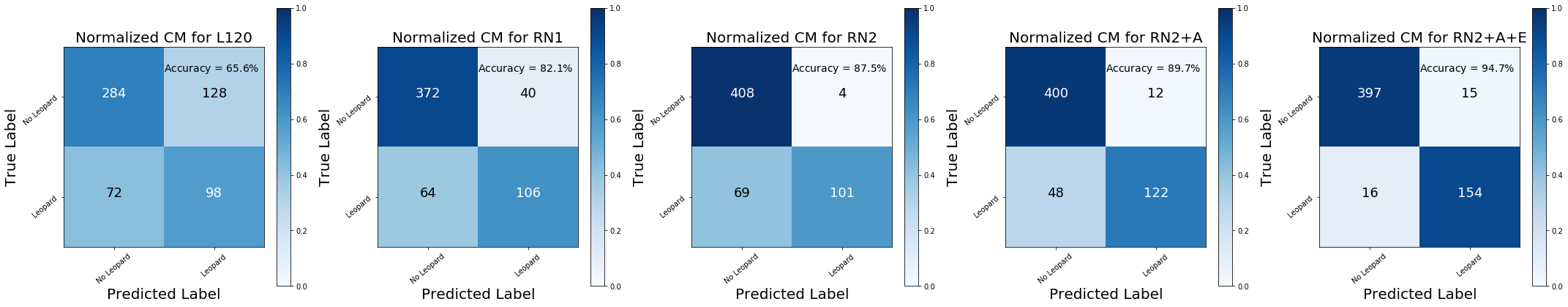}
  \caption{Normalized confusion matrices for each algorithm. }
  \label{fig:accuracy}
\end{figure}

Finally, we leverage the natural grouping of camera trap images into in small temporally contiguous segments we refer to as ``bursts''. We build on Spark's primitives to implement more complex inter-node operations like grouping images by keys. These bursts originate from camera traps that take a series of images until no motion is detected. We extract this burst information from image metadata, and use it to inform our predictions. This effectively turns a single image classifier into a video classifier. We note that labels are often the same for all frames in camera bursts. By averaging the predictions of each image in a burst, we dramatically reduce the variance of the classifier and allow the algorithm to use easy classifications to inform neighboring difficult classifications. This substantially reduces the number of false negatives, and improves overall accuracy. We refer to this algorithm as ``RN2+A+E'' for the base algorithm with dataset augmentation and sequence ensembling. We compare these models to logistic regression trained on the raw pixels of the image after scaling to $120 \times 120$ pixels. We refer to this baseline model as ``LR120''. Figures \ref{fig:accuracy} and \ref{fig:roc} show the dramatic performance increases from deep featurization, augmentation, and ensembling.

\section{Future Work}

MMLSpark is a growing software ecosystem with several large ongoing areas of work. In the future, we plan to extend the CNTK Java bindings to allow for training in Java. This will allow users to construct CNTK computation graphs in Java and Scala, and will provide a way to leverage the parallelism of the spark cluster for CNTK model training in addition to evaluation. 

Furthermore, we plan to implement a recent advance in low communication parallel stochastic gradient descent called SymSGD \cite{symsgd}. This will dramatically reduce communication overhead during parallel training of networks. In addition this improve the scalability of parallel training systems as SymSGD's method of gradient combination approximates a sequential computation. This contrasts other methods like AllReduce, ParameterServer, or Hogwild! that principally benefit from reducing the noise of the gradient. These methods see rapidly diminishing marginal returns as the number of workers increases \cite{symsgd}.

\section{Conclusions}

We have created an architecture for embedding arbitrary CNTK computation graphs and OpenCV computations in Spark computation graphs. This architecture inter-operates with the majority of the Spark ecosystem and enables fault tolerant deep learning at massive scales. We also successfully translate this and the rest of the MMLSpark ecosystem into PySpark and SparkR, enabling scalable deep learning in a variety of languages. We leverage this framework to create a state-of-the art snow leopard detection system for the Snow Leopard Trust. This work lays the groundwork for future deep learning extensions to the Spark ecosystem and enables convenient cluster based solutions for a broader class of machine learning problems. 


\acks{The authors would like to thank the Snow Leopard Trust, specifically Rhetick Sengupta and Koustubh Sharma, for their continued cooperation, inspiration, and dataset. Thanks to Patrick Buehler for connecting the MMLSpark team with the Snow Leopard Trust and for creating an initial POC of a snow leopard detector. In addition, we thank the CNTK team, specifically Mark Hillebrand, Wolfgang Manousek, Zhou Wang, and Liqun Fu for their continued support around SWIG and the CNTK build system. Finally, without the hard work and resources of the Azure Machine Learning team and Microsoft, this project would not have been possible. }

\vskip 0.2in
\bibliography{sample}

\end{document}